\documentclass[fleqn,usenatbib]{mnras}

\usepackage{newtxtext,newtxmath}

\usepackage[T1]{fontenc}

\DeclareRobustCommand{\VAN}[3]{#2}
\let\VANthebibliography\thebibliography
\def\thebibliography{\DeclareRobustCommand{\VAN}[3]{##3}\VANthebibliography}

\usepackage{graphicx}	
\usepackage{amsmath}	

\title[Detection of Forbidden Lines from the AGN BLR?]{The First Detection of Forbidden Emission Lines at the Outskirts of the AGN Broad Line Region?}
\author[K. F. Heckler et al.]{Kelly F. Heckler,$^{1,2}$\thanks{E-mail: kelly.heckler@acad.ufsm.br}
Rogemar A. Riffel,$^{1}$\thanks{E-mail: rogemar@ufsm.br}
Daniel Marsango,$^{1}$
Tiago V. Ricci,$^{3}$
Angela C. Krabbe,$^{2}$
\newauthor
and Oli L. Dors,$^{4}$
\\
$^{1}$Departamento de F\'isica, CCNE, Universidade Federal de Santa Maria, 97105-900 Santa Maria, RS, Brazil\\
$^{2}$Departamento de Astronomia, IAG, Universidade de São Paulo, 05508-900 São Paulo, SP, Brazil\\
$^{3}$Universidade Federal da Fronteira Sul, Campus Cerro Largo, 97900-000 RS, Brazil\\
$^{4}$Universidade do Vale do Para\'iba, Av. Shishima Hifumi, 2911, Cep
12244-000, S\~ao Jos\'e dos Campos, SP, Brazil 
}

\date{Accepted XXX. Received YYY; in original form ZZZ}

\pubyear{\the\year{}}

\begin{document}
\label{firstpage}
\pagerange{\pageref{firstpage}--\pageref{lastpage}}
\maketitle

\begin{abstract}
Double-peaked (DP) broad emission line profiles in active galactic nuclei (AGNs) are often interpreted as signatures of a rotating disk-like structures in the broad-line region (BLR) and are commonly observed in low luminosity AGNs using recombination lines. We use optical spectroscopy to investigate the origin of double-peaked broad emission line profiles observed not only in hydrogen recombination lines but also in forbidden transitions in the LINER galaxy IC~1459. 
We detected DP emission in all strong optical lines, except for the [S~\textsc{ii}] doublet, which has the lowest critical density among all the lines. We successfully fitted the DP broad profiles using a disk-like BLR model, assuming a circular accretion disk with an inclination of $\sim 35^\circ$ and internal turbulence of $\sim 500$~km\,s$^{-1}$, confined within a maximum radius of  $9.6^{+4.8}_{-1.1}$ light-years. We estimate a full width at half maximum of the DP profiles of FWHM~$\sim 3300$~km\,s$^{-1}$. Our results provide new insights into the structure of the BLR, indicating that forbidden emission lines can be produced in lower-density regions near the outskirts of the BLR.
\end{abstract}




\section{Introduction}

Active Galactic Nuclei (AGNs) are powered by the accretion of matter onto a supermassive black hole (SMBH) located at the center of galaxies. The innermost structure of AGN hosts the unresolved broad-line region (BLR), which consists of gas clouds with very high electron density \citep[e.g. $N_{\rm e} \: > \: 10^{8}$,][]{2006A&A...447..157Nagao} photoionized by the intense radiation emitted by the accretion disc. The interaction between the radiation and the gas results in broad permitted emission lines, which are characteristic of the BLR. The profiles of these broad emission lines reveal the complex dynamics of the BLR gas clouds, exhibiting features such as asymmetries, double-peaked emissions, and a wide range of line widths \citep[][and references therein]{2000ARA&A..38..521Sulentic}.

Double-peaked broad emission line profiles are relatively rare in the literature, as only a small fraction (3\% -- 10\%) of the general AGN population exhibit such features \citep{1994ApJS...90....1Eracleous, 2003AJ....126.1720Strateva, 2024ApJ...961..172Ward}. These profiles are predominantly observed in low-luminosity AGNs \citep{1994ApJS...90....1Eracleous, 2000ApJ...541..120Ho, 2014MNRAS.438.3340Elitzur}. Such profiles can be attributed to emission from a relativistic Keplerian disc, as first proposed by \citet{1989ApJ...339..742Chen} and \citet{1989ApJ...344..115Chen} based on H$\alpha$ observations of the Type-I AGN Arp 102B. To explain the asymmetries observed in double-peaked profiles, various disc models have been proposed, including elliptical \citep{1995ApJ...438..610Eracleous, 1997ApJ...489...87StorchiBergmann}, spiral-like \citep{1999ASPC..175..189Gilbert}, warped \citep{2008MNRAS.389..213Wu}, and rotating hot spots \citep{1997ApJ...485..570Newman}. Among these, circular, elliptical, and spiral-like disc models are the most commonly used. The elliptical and spiral-like models are particularly favored in cases where the double-peaked profile exhibits temporal variability and/or when the red peak is more prominent than the blue one \citep{2003ApJ...598..956Storchi-Bergmann, 2003ApJ...599..886Eracleous, 2003AJ....126.1720Strateva, 2012ApJ...748..145SSchimoia, 2017MNRAS.472.2170Schimoia, 2025arXiv250321994Ochmann}.

Double-peaked profiles are typically observed in the optical region using hydrogen recombination lines, although similar features have also been reported in the ultraviolet \citep{2003AJ....126.1720Strateva, 2019ApJ...877...33Zhang, 2022MNRAS.516.5775Bianchi} and infrared wavelengths \citep[e.g.,][]{2023ApJ...953L...3DiasDosSantos, 2025arXiv250321994Ochmann}. Recent  near-infrared studies of Seyfert galaxies suggest that the double-peak emission originates from different locations within the BLR. For instance, \citet{2023ApJ...953L...3DiasDosSantos} reported the detection of double-peaked O\,{\sc i}\,$\lambda11297$ in III\,Zw\,002, a nearby ($z=0.0898$)  type 1 AGN, which traces the emission from the outer regions of the BLR. Conversely, \citet{2025arXiv250321994Ochmann} found that the double-peaked Ca\,{\sc ii} triplet observed in NGC~4593 (type 1 AGN, $z=0.008312$) is well described by a mildly eccentric, low-inclination relativistic line-emitting disk with minimal internal turbulence. This double-peaked emission is associated with a region characterized by gas densities of the order of $n_\mathrm{H} \sim 10^{11.5}$ cm$^{-3}$ \citep{2007ApJ...663..781Matsuoka}.

Understanding the physical conditions of the emitting region is essential, and one key parameter is the electron density. Although it does not directly influence the line profiles, electron density determines which emission lines are present. The BLR is a high-density environment, with electron densities on the order of $\sim 10^{10}$ cm$^{-3}$ \citep{1979RvMP...51..715Davidson, 1989ApJ...347..656Ferland}, leading to the observation of permitted and semi-forbidden emission lines \citep{2000ARA&A..38..521Sulentic}. At such high densities, collisional de-excitation dominates, effectively suppressing forbidden transitions. Consequently, only permitted lines are observed in these regions. In contrast, in areas with lower electron densities, such as the narrow-line region \citep[$N_{\rm e} < 10^{4} \rm cm^{-3}$, e.g. ][]{2020MNRAS.492..468Dors}, the probability of collisional de-excitation is significantly reduced. This allows metastable states to decay radiatively, producing the forbidden lines.

Forbidden lines typically have low critical densities, as summarized in Table~\ref{tab:dens} for commonly observed strong optical lines in AGNs. In this work, we report the first detection of double-peaked forbidden emission lines originating from the outskirts of the BLR in the low-luminosity AGN (log L$_\mathrm{bol} = 43.09$ erg s$^{-1}$) of IC~1459 (z $= 0.006011$), classified as LINER using diagnostic diagrams \citep{Ricci2014_II, 2023MNRAS.522.2207Ricci}. 
The structure of this paper is as follows: in Section 2, we describe the observations, data processing and the method used to subtract the stellar population contribution from the observed spectra. In Section 3, we present the results and discuss the BLR disc model adopted to reproduce the observed double-peaked emission line profiles. Finally, in Section 4, we summarize our main conclusions and discuss their implications in the context of AGN broad-line region studies.

\begin{table}	\centering
	\caption{Critical densities of optical forbidden emission lines as calculated with PyNeb code \citep{luridiana15}. The ionization potential (IP) was obtained from \citet{NIST_ASD}. }
	\label{tab:dens}
	\begin{tabular}{lccc} 
		\hline
Line & Transition & $N_c$ (cm$^{-3}$) & IP (eV)\\
\hline
$\left[ \text{O}\, \text{III} \right] \lambda 4959$ & $4 \rightarrow 2$ & $7.83\times10^5$ & 35.12  \\
$\left[ \text{O}\, \text{III} \right] \lambda 5007$ & $4 \rightarrow 3$ & $7.83\times10^5$ & 35.12 \\
$\left[ \text{O}\, \text{I} \right] \lambda 6300$ & $4 \rightarrow 1$ & $1.24\times10^6$ & 13.62\\
$\left[ \text{O}\, \text{I} \right] \lambda 6363$ & $4 \rightarrow 2$ & $1.24\times10^6$ & 13.62 \\
$\left[ \text{N}\, \text{II} \right] \lambda 6548$ & $4 \rightarrow 2$ & $1.04\times10^5$ & 14.53  \\
$\left[ \text{N}\, \text{II} \right] \lambda 6583$ & $4 \rightarrow 3$ & $1.04\times10^5$ & 14.53  \\
$\left[ \text{S}\, \text{II} \right] \lambda 6717$ & $3 \rightarrow 1$ & $1.92\times10^3$ & 10.36  \\
$\left[ \text{S}\, \text{II} \right] \lambda 6731$ & $2 \rightarrow 1$ & $5.07\times10^3$ & 10.36  \\
\hline
        \end{tabular}

\end{table}

\section{Data and spectral fitting}

\subsection{Observations and Data Processing}

IC 1459 was observed in August 2008 with the Gemini South Telescope under the GS-2008B-Q-21 programme. The observations were made with the Gemini Multi-Object Spectrograph \citep[GMOS;][]{AllingtonSmith_2002} with the integral field unit (IFU) in one-slit mode. B600 grating centered in 5650 \AA\, was employed, providing a spectral coverage from 4228 \AA\, to 7121 \AA, and has a spectral resolution of 1.8 \AA, as estimated from the O I $\lambda$5577 sky line \citep{Ricci2014_I}. The spatial resolution is approximately 0.7 arcsec (corresponding to a projected physical scale of 91~pc), as estimated by measuring the full width at half maximum (FWHM) of the flux distribution of field stars in the acquisition image of the observed object. A large scale image of IC 1459 is presented in \citet{Ricci2014_I}, with the GMOS field-of-view overplotted. 

We used the IRAF package for the Gemini telescopes to perform data reduction, following the standard procedure: bias and flat-field corrections, cosmic ray removal via the LACOS algorithm \citep{van_Dokkum_2001}, wavelength and flux calibration, and the construction of the data cube with a pixel scale of 0.05 arcseconds \citep{Ricci2014_I}. 
Finally, high- and low-frequency noise along both spatial and spectral dimensions was removed from the data cube using techniques outlined in \citet{menezes2019}, with additional details in \citet{Ricci2014_I}. For further information regarding the data reduction and processing methods, see \citet{Ricci2014_I} and \citet{menezes2019}.

\subsection{Subtraction of the stellar population contribution}

In order to study in detail the observed gas emission for the IC 1459 galaxy, it is necessary to isolate the gas component from the observed spectrum. For this purpose, we used of the Penalized Pixel-Fitting ({\sc ppxf}) method \citep{Cappellari2017_ppxf, 2023MNRAS.526.3273Cappellari_ppxf} to model the stellar absorption features, which were subsequently subtracted from the observed spectrum.
We selected a set of stellar spectra with a spectral resolution similar to that of our GMOS data: the MILES-HC library \citep[2.5 \AA;][]{Westfall_2019}. 
In the fitting process, the stellar templates were convolved with the line-of-sight velocity distribution (LOSVD) represented by a Gauss–Hermite series up to the fourth moment. 
We used a 12th-order multiplicative Legendre polynomial to correct the continuum shape, which adequately reproduced the observed spectrum without the need for an additive polynomial or sky contribution.
As our aim was solely to remove the stellar contribution from the observed spectrum, without delving into the properties of the stellar populations, we find that the chosen library successfully reproduces the observed stellar features.

\section{Results and discussion}

\begin{figure*}
    \centering
    \includegraphics[width=\linewidth]{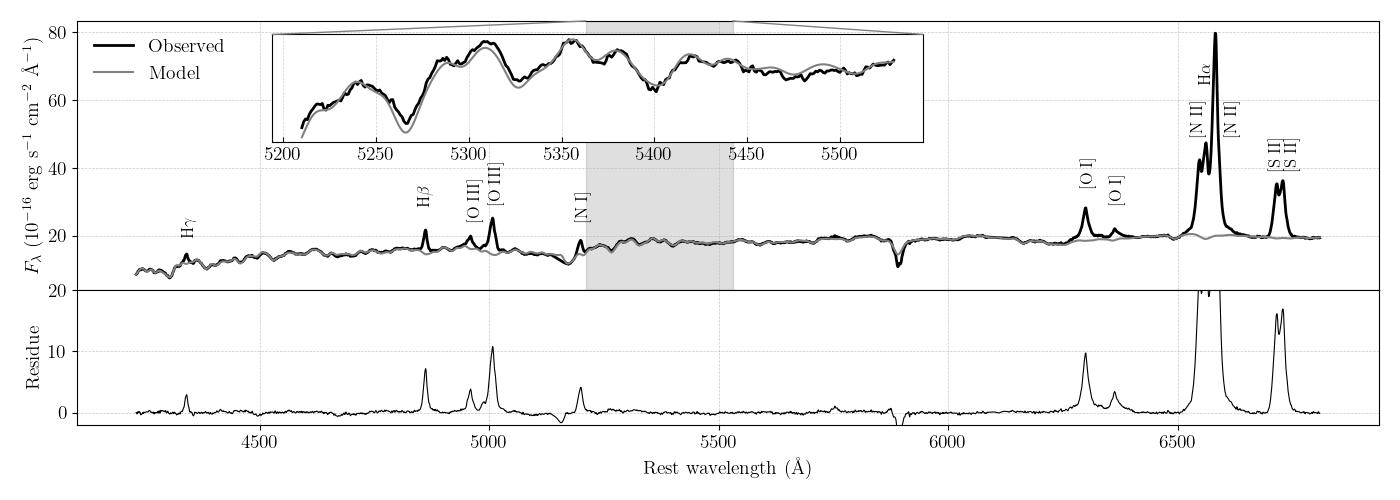}
    \caption{Stellar population model obtained using the {\sc ppxf} fit of the integrated nuclear spectrum. The black line shows the observed spectrum and the \textbf{gray} line represents the stellar population model. A zoomed-in view of a selected spectral region is included to illustrate the fit quality. The bottom panel shows the residuals of the fit, which are overall small expect for the locations of the emission lines. }
    \label{fig:nuclear_spec}
\end{figure*}

By visually inspecting the spectrum, we identified that the emission lines exhibit a broad base with an unresolved flux distribution concentrated at the galaxy's nucleus.  An analysis of the extended narrow-line emission is presented in \citet{Ricci2015_III}. \citet{2023MNRAS.522.2207Ricci} reported a broad component in H$\alpha$, H$\beta$ and in the [O\,{\sc i}]$\lambda\lambda$ 6300, 6363 doublet, with a FWHM $\sim$ 3500 km/s for the Balmer lines and $\sim$ 2700 km/s for the [O\,{\sc i}] lines. They speculated that this component may be emerging from the outermost region of the BLR. In this work, we further investigate this scenario in detail. Here, we focus on the origin of the unresolved nuclear broad component. For this purpose, we extracted a spectrum within a circular aperture with a radius of $r = \text{0.35}$ arcsec, centred at the position of the galaxy nucleus, defined as the location of the continuum peak. This aperture approximately corresponds to the angular resolution of the observations, thus representing the nuclear emission of the galaxy.  In Figure \ref{fig:nuclear_spec}, we show the resulting observed spectrum in black, along with the best-fit model of the stellar population contribution, obtained using the {\sc ppxf} method, shown in gray. 
In addition, we show a zoomed-in region free of emission lines, demonstrating that the stellar population model satisfactorily reproduces the observed continuum and absorption spectrum. Thus, by subtracting the stellar population model from the observed spectrum, we obtain a spectrum free of the stellar component, allowing us to investigate the gas emission.

\begin{figure*}
    \centering
    \includegraphics[width=\textwidth]{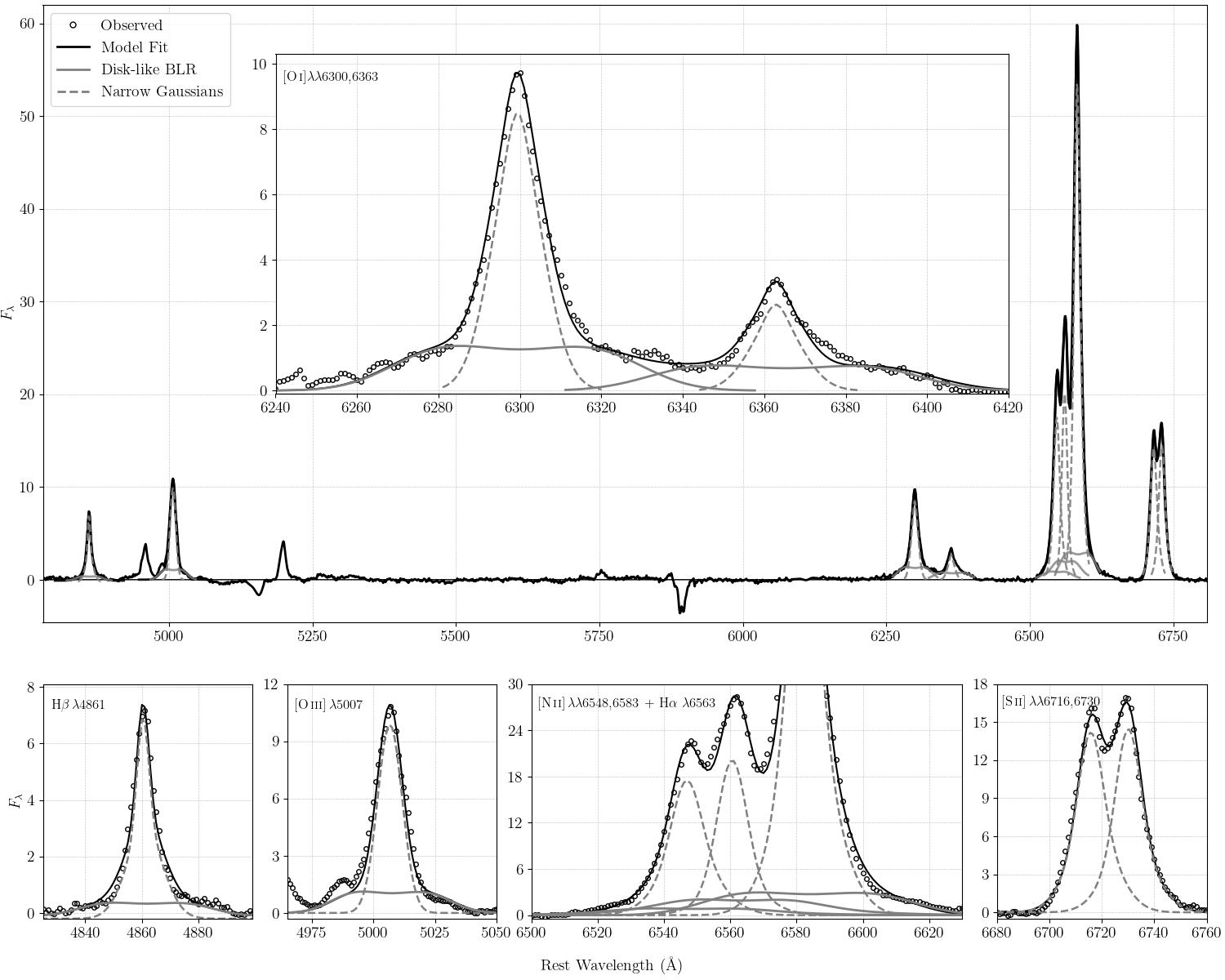}
   \caption{Nuclear spectrum of IC~1459 after subtraction of the stellar population contribution (black circles). The insets show the emission line profiles, along with the best-fit model (black line). The disk-like BLR component is shown in gray line, while the narrow components of the emission lines are shown in gray dotted line. }
   \label{fig:model}
\end{figure*}

In Figure~\ref{fig:model}, we show the stellar-population-subtracted spectrum as black circles, with insets highlighting different emission lines. Broad component emission is detected in the [O\,{\sc iii}]\,$\lambda \lambda4959,5007$, [O\,{\sc i}]\,$\lambda \lambda6300,6364$ and [N\,{\sc ii}]\,$\lambda \lambda6548,6583$ forbidden emission lines, as well as in H$\beta$ and H$\alpha$. 
In addition, Figure~\ref{fig:plot_oi} presents the residual emission in the [O\,\textsc{i}]$\lambda\lambda6300,6364$ doublet region, obtained after subtracting the stellar continuum and the narrow-line components, allowing the DP broad components to be clearly identified.
On the other hand, the $\left[ \text{S}\, \text{II} \right] \lambda\lambda 6717,6731$ emission lines do not show broad components. Broad components in forbbiden emission lines of AGN hosts are usually associated with outflows, which are often observed blueshifted relative to the narrow components \citep[e.g.][]{Crenshaw10,Wylezalek20,Santoro20}. However, the broad lines detected in IC~1459 exhibit profiles that differ from those expected for outflowing gas. In particular, they show a flattened profile on both the redshifted and blueshifted sides relative to the narrow line, closely resembling the double-peaked broad-line profiles observed in the AGN BLR \citep[e.g.][]{storchi-bergamann95,Strateva03,Eracleous09,Marsango24}. 

Although such double-peaked profiles could, in principle, be interpreted as signatures of BLR outflows, previous studies have shown that BLR-driven winds are typically asymmetric and strongly blueshifted, with velocities in the range of $4\,000 - 11\,000$ km s$^{-1}$ \citep{2011ApJ...738...85Wang}. Following the interpretation proposed by \citet{2011ApJ...738...85Wang}, the presence of a symmetric double-peaked profile with FWHM of $\sim 3\,300$ km s$^{-1}$ is therefore unlikely to be attributed to outflows, since such models tend to produce asymmetric, blueshifted line profiles. The simultaneous detection of prominent blue- and redshifted components disfavors an outflow-dominated scenario and instead points to an origin in the rotating disk of the BLR, where orbital motions naturally give rise to symmetric double-peaked emission-line profiles.

We also consider the possibility that the observed broad emission may arise from a binary black hole system, which can produce complex emission-line profiles. However, despite IC~1459 showing evidence of a past merger \citep{1988ApJ...327L..55Franx, 2003MNRAS.346..327Tingay}, the likelihood that the broad emission is associated with a binary black hole is low. This interpretation is further disfavored by the presence of unresolved hard X-ray emission associated with a single AGN \citep{1997ApJ...482..133Matsumoto}, as well as a compact radio source with a diameter smaller than 0.03 arcsec \citep{1994MNRAS.269..928Slee}.

Thus, a possible explanation for the double-peaked broad-line profiles in IC~1459 is that they originate in the outer, lower-density ($\geq 10^5 {\rm cm}^{-3}$) regions of the AGN BLR. The non-detection of the broad component in the [S\,{\sc ii}] doublet is naturally explained by their much lower critical densities (Table~\ref{tab:dens}). In addition, among the forbidden lines, the double-peaked line component is clearly observed in the [O\,{\sc i}] lines, which have the highest critical densities. 
This picture is consistent with a stratified BLR structure, in which the inner regions reach electron densities of $10^8 - 10^{11} {\rm cm}^{-3}$, while the density decreases outward \citep{2004ApJ...606..749Korista, 2015MNRAS.453.3662Goad}.
In a previous study, \citet{2021MNRAS.504.5087Mulcahey} investigated the origin of the gas emission in IC 1459 using MUSE (Multi-Unit Spectroscopic Explorer) data from the Very Large Telescope (VLT), but did not detect double-peaked emission in any of the emission lines. This non-detection could be attributed to the lower spatial resolution of MUSE compared to GMOS.

To further explore the origin of the double-peaked line profiles in IC~1459, we fit the observed profiles using a disk-like BLR model, based on the model originally proposed by \citet{1989ApJ...339..742Chen} to reproduce the line profiles of Arp\,102B. This model has been successfully applied to reproduce double-peaked profiles in nearby AGN \citep{ricci,2017MNRAS.472.2170Schimoia,Storchi-Bergmann_2017}.  In brief, to model the surface emissivity of this region, we adopt a radial dependence of the emissivity between the inner and outer radii of the emitting region, $\xi_1$ and $\xi_2$, as given by \citep{2012ApJ...748..145SSchimoia}:
\begin{equation}
\epsilon_{R}(\xi) =
\begin{cases}
\xi^{-q_1}, & \xi_1<\xi<\xi_q \\
\xi_q^{-(q_1-q_2)}\xi_q^{-q_1}, & \xi_q<\xi<\xi_2
\end{cases}
\end{equation}
\noindent where $\xi_q$ corresponds to the radius of maximum emissivity, $q_1$ is the emissivity index for the region between $\xi_1$ and $\xi_q$, while $q_2$ corresponds to the index for the region between $\xi_q$ and $\xi_2$. In this model, the double-peaked emission line originates at the surface of the disk, whose axis is inclined by an angle $i$ relative to the plane of the sky. The line broadening is attributed to the turbulent motion of the gas in the emitting cells, represented by a Gaussian profile with velocity dispersion $\sigma$ (in km s$^{-1}$). 

We start by fitting the [O\,{\sc i}]\,$\lambda6300$ emission line because it is not blended with other lines and exhibits the most prominent double-peaked profile.
 The parameters that provided the best fit are: $i = 35^\circ{}^{+12.08}_{-14.69}$, 
$\sigma = 500^{+250}_{-199}$ km s$^{-1}$, 
$\xi_1 = 10\,000^{+5\,000}_{-4\,184}\,R_g$, 
$\xi_2 = 28\,000^{+14\,000}_{-3\,143}\,R_g$, 
$\xi_q = 15\,000^{+7\,500}_{-6\,888}\,R_g$, 
$q_1 = -1.00^{+1.00}_{-0.11}$, 
and 
$q_2 = 3^{+1.26}_{-1.50}$\footnote{The uncertainties in the parameters were obtained by varying each parameter separately around the best-fit value while keeping all other parameters fixed.}
, where $R_g$ is the gravitational radius, defined as:
\begin{equation}
    r_g = \frac{2G M_{\bullet}}{c^2}
\end{equation}
\noindent where $G$ is the gravitational constant, $M_{\bullet}$ is the mass of the SMBH, and $c$ is the speed of light. Then, the same fitting parameters were used to reproduce the remaining emission lines (H$\beta$\,$\lambda$4861, [O~\textsc{iii}]~$\lambda$5007, [N~\textsc{ii}]~$\lambda$6548, H$\alpha$\,$\lambda6563$ and [N~\textsc{ii}]~$\lambda$6583).

The insets in Fig.~\ref{fig:model} show the observed line profiles along with the models (see Table \ref{tab:line_params}). In addition to the disk-like BLR model, we fit the narrow components of the line profiles using a single Gaussian per line. As can be seen, the observed profiles are very well reproduced by the composite model, including the emission from both the BLR and the NLR.
In the [O\,{\sc iii}]\,$\lambda 5007$ panel, a blue wing is visible that is not reproduced by either the gaseous disc or the BLR disc model. A plausible explanation for this feature is outflowing gas, but a detailed discussion of the origin of this feature is beyond the scope of this work. 
Using the modelled double-peak profile, we estimate a full width at half maximum of FWHM$\approx$3\,300 km\,s$^{-1}$, which is similar to the FWHM obtained by \citet{2025arXiv250321994Ochmann} for the double-peaked Ca\,{\sc ii} triplet and O\,{\sc i} ($\sim 3\,700$ km s$^{-1}$) and smaller than the typical values for double-peaked H recombination line profiles, which are in the range 5\,000 -- 23\,000 km\,s$^{-1}$ \citep{2017ApJ...835..236StorchiBergmann, 2023ApJ...953L...3DiasDosSantos, 2025ApJ...987...14Wu}.
Thus, this work presents for the first time the detection of disk-like BLR emission in forbidden lines, originating in regions with densities higher than $\sim10^{5}$ cm$^{-3}$, which corresponds to the critical density of the [N\,{\sc ii}] lines — the lowest among the detected double-peaked lines. 

We estimate the gravitational radius of the SMBH to be $R_g = 0.125$ light-days. Assuming that the emitting region extends from $10\,000\,R_g$ to $28\,000\,R_g$; this corresponds to a radial range between $R_{\mathrm{in}} = 3.4^{+1.7}_{-1.4}$ and $R_{\mathrm{out}} = 9.6^{+4.8}_{-1.1}$ light-years. The total extent of the emitting region is therefore approximately $6.2^{+6.2}_{-2.8}$ light-years. This value is larger than the typical physical size estimated for BLR in other AGNs \citep{2003AJ....126.1720Strateva, 2025arXiv250321994Ochmann, 2023ApJ...953L...3DiasDosSantos, Gravity_2024A&A...690A..76G, gravity_amorin, 2024Natur.627..281Abuter}. 
Given the specific conditions required for the emission of forbidden lines, it is plausible that the emitting region in this case is intrinsically more extended than the regions traced by recombination lines, occupying a region of lower gas density compared to the locations where the recombination lines are produced. This is consistent with recent BLR models that assume the gas density decreases from the nucleus following a power-law \citep{Rosborough24}. 

Finally, the possibility of an intermediate emission region located between the BLR and the NLR cannot be ruled out. This region, known as the inner NLR  \citep[INLR; ][]{2016A&A...586A..48Balmaverde}, exhibits gas emission properties characterized by intermediate line FWHM values, arranged in a preferential plane -- thus explaining the successful kinematic reproduction by the model of \citet{1989ApJ...339..742Chen}. Moreover, it appears to be more distant and spatially more extended than the region traditionally attributed to the BLR, possibly forming a gas ring in the outermost BLR region. This interpretation would be consistent with our estimate for the BLR radius. 
In addition, outflows in the BLR can also produce complex and DP broad-line profiles, which can be reproduced by models of clumpy winds illuminated by the AGN continuum, naturally producing asymmetric and DP BLR line profiles \citep[e.g.][]{Kollatschny13,Elitzur14,Shin17,Matthews2020}. In this context, future studies using high-resolution spectroscopy (e.g., adaptive-optics-assisted integral-field observations) and time-domain monitoring will be key to unveiling the origin of the broad forbidden emission-line profiles in IC~1459 reported here.

\section{Conclusions}

In this work, we investigated the ionized gas emission in the nuclear region ($\sim$91~pc) of the LINER galaxy IC~1459 using Gemini GMOS-IFU data. We report, for the first time, the detection of double-peaked broad profiles in several optical emission lines, including both permitted and forbidden transitions, produced in the outskits of the BLR. Double-peaked emission was detected in all optical lines observed (H$\beta$~$\lambda4861$, [O~\textsc{iii}]~$\lambda5007$, [O~\textsc{i}]~$\lambda6300$, [N~\textsc{ii}]~$\lambda6548$, H$\alpha$~$\lambda6563$, and [N~\textsc{ii}]~$\lambda6583$), with the sole exception of the [S~\textsc{ii}] doublet.  The double-peaked profiles are well reproduced by a disk-like BLR model, which corresponds to a circular accretion disk with an inclination of $\sim 35^\circ$, internal turbulence of $\sim 500$~km\,s$^{-1}$, and a maximum radius of $28\,000\,R_g$, corresponding to $R_{\mathrm{out}} = 9.6^{+4.8}_{-1.1}$ light-years. 
This implies that the density of the emitting region must be larger than the [S\,\textsc{ii}] critical density, 
i.e., $n_{\mathrm{e}} \gtrsim 5\times10^3\:{\rm cm^{-1}}$.

These findings provide important insights into the structure and physical conditions of the BLR in low-luminosity AGNs, showing that forbidden emission lines can be produced by lower-density gas in the outskirts of the BLR. Meanwhile, the innermost BLR is typically traced by recombination lines, which arise from higher-density gas.

\section*{Acknowledgements}
The authors are grateful to the anonymous referee for the valuable comments and constructive suggestions. 
Based on observations obtained at the Gemini Observatory, which is operated by the Association of Universities for Research in Astronomy, Inc., under a cooperative agreement with the NSF on behalf of the Gemini partnership: the National Science Foundation (United States), National Research Council (Canada), CONICYT (Chile), Ministerio de Ciencia, Tecnolog\'{i}a e Innovaci\'{o}n Productiva (Argentina), Minist\'{e}rio da Ci\^{e}ncia, Tecnologia e Inova\c{c}\~{a}o (Brazil), and Korea Astronomy and Space Science Institute (Republic of Korea). 
This research has made use of NASA's Astrophysics Data System Bibliographic Services. This research has made use of the NASA/IPAC Extragalactic Database (NED), which is operated by the Jet Propulsion Laboratory, California Institute of Technology, under contract with the National Aeronautics and Space Administration. ChatGPT (GPT-4.5) was used to assist with code debugging and minor language editing.
KFH thanks the financial support from the Conselho Nacional de Desenvolvimento Científico e Tecnológico (CNPq; Project 140760/2020-2) and the Fundação de Amparo à Pesquisa do Estado de São Paulo (FAPESP), Brasil (Process number 2024/22775-9). 
RAR acknowledges the support from the Conselho Nacional de Desenvolvimento Científico e Tecnológico (CNPq; Projects 303450/2022-3, 403398/2023-1, and 441722/2023-7) and the Coordenação de Aperfeiçoamento de Pessoal de Nível Superior (CAPES; Project 88887.894973/2023-00). DM acknowledges support from CAPES (Finance code 001). 
TVR thanks CNPq for support under grant 304584/2022-3.
A.C.K. thanks the FAPESP for the support grant 2024/05467-9 and the CNPq grant 315566/2023-0.

\section*{Data Availability}

The data used in this work are publicly available online via the GEMINI archive https://archive.gemini.edu/searchform, under the program code GS-2008A-Q-51.


\bibliographystyle{mnras}
\bibliography{references} 




\appendix

\section{Residual double-peaked broad profile}

\begin{figure}
    \centering
    \includegraphics[width=\linewidth]{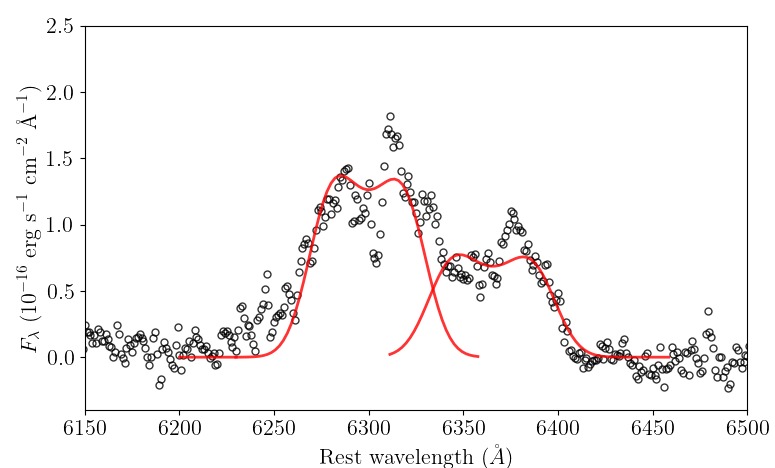}
    \caption{Residual emission after subtraction of stellar continuum and narrow line components for the [O\,{\sc i}]\,$\lambda \lambda 6300, 6364$ doublet region is represented by the black circles. The red lines represent the BLR disc model. }
    \label{fig:plot_oi}
\end{figure}

In Figure \ref{fig:plot_oi}, we present the residual emission obtained after subtracting the stellar continuum and the narrow-line components, in order to clearly highlight the broad component modeled with the BLR disk profile.

Table \ref{tab:line_params} lists the best-fitting parameters derived for both the narrow and BLR components. The narrow emission-line profiles were modeled using Gaussian functions, whereas the broad component was fitted with the BLR disk model described by \citet{1989ApJ...339..742Chen}.

\begin{table}
\centering
\caption{Parameters from the BLR and narrow emission line models from Figure \ref{fig:model}. Narrow-line measurements were performed assuming Gaussian profiles. For the BLR components, the line flux was obtained from the integrated BLR emission, the centroid velocity from the first moment of the flux distribution relative to the rest wavelength, and the FWHM from the standard full-width at half-maximum definition applied to the integrated profile. 
}
\label{tab:line_params}
\begin{tabular}{r c c c c}
\hline
Line & Flux & $v_{\rm c}$ &  FWHM \\
    &  (10$^{-15}$ & (km\,s$^{-1}$) & (km\,s$^{-1}$) \\
        & $\text{erg s}^{-1} \text{cm}^{-2}$) & & & \\
\hline
H$\beta_{\,\mathrm{BLR}}$                              &  16 &   17 & 2801 \\
H$\beta_{\,\mathrm{Narrow}}$                          &  64 &  -13 &  684 \\ 
$\left[ \text{O}\,\text{\sc iii} \right]\lambda5007_{\,\mathrm{BLR}}$     &  57 &   4  & 3031 \\ 
$\left[ \text{O}\,\text{\sc iii} \right]\lambda5007_{\,\mathrm{Narrow}}$ & 124 &  -21 &  706 \\
$\left[ \text{O}\,\text{\sc i} \right]\lambda6300_{\,\mathrm{BLR}}$       &  87 &  -38 & 3325 \\
$\left[ \text{O}\,\text{\sc i} \right]\lambda6300_{\,\mathrm{Narrow}}$   & 123 &  -27 &  689 \\
$\left[ \text{O}\,\text{\sc i} \right]\lambda6364_{\,\mathrm{BLR}}$       &  53 &   87 & 3260 \\
$\left[ \text{O}\,\text{\sc i} \right]\lambda6364_{\,\mathrm{Narrow}}$   &  37 &   -5 &  681 \\
H$\alpha_{\,\mathrm{BLR}}$                            & 129 &   16 & 2644 \\
H$\alpha_{\,\mathrm{Narrow}}$                        & 262 & -100 &  659 \\
$\left[ \text{N}\,\text{\sc ii} \right]\lambda6583_{\,\mathrm{BLR}}$      & 197 &   22 & 3031 \\
$\left[ \text{N}\,\text{\sc ii} \right]\lambda6583_{\,\mathrm{Narrow}}$  & 824 &  -13 &  777 \\
$\left[ \text{S}\,\text{\sc ii} \right]\lambda6716_{\,\mathrm{Narrow}}$  & 228 &  -11 &  791 \\
$\left[ \text{S}\,\text{\sc ii} \right]\lambda6731_{\,\mathrm{Narrow}}$  & 223 &  -14 &  764 \\
\hline
\end{tabular}
\end{table}



\bsp	
\label{lastpage}
\end{document}